\documentclass[twocolumn,english,aps,pre]{revtex4}

\usepackage{amsmath}
\usepackage{graphicx}
\usepackage{amssymb}
\usepackage{babel}
\usepackage{color}
\renewcommand{\vec}[1]{\mathbf{#1} }

\date{April 5th 2011}


\begin{document}

\title{Computer simulations of colloidal transport on a patterned magnetic substrate}

\author{Andrea Fortini}

\author{Matthias Schmidt}

\affiliation{ Theoretische Physik II, Physikalisches Institut, Universit\"at Bayreuth, Universit\"atsstra{\ss}e 30, D-95447 Bayreuth,
Germany}

\begin{abstract}
We study  the transport of paramagnetic colloidal particles on a patterned magnetic substrate with kinetic Monte Carlo  and Brownian dynamics computer simulations. The planar substrate is decorated with point dipoles in either parallel or zigzag stripe arrangements and exposed to an additional external magnetic field that oscillates in time.
For the case of parallel stripes we find that the magnitude and direction of the particle current  is controlled by the tilt angle of the external magnetic field. The effect is reliably obtained in a wide range of ratios between  temperature and magnetic permeability. Particle transport is achieved only when the period of oscillation of the external field is greater than a critical value.
For the case of zigzag stripes a current is obtained using an oscillating external field normal to the substrate. 
In this case,  transport is only possible in the vertex of the zigzag, giving rise to a narrow stream of particles.
The magnitude and direction of the particle current  are found to be controlled by a combination of the zigzag angle  and the distance of the colloids from the substrate.
Metropolis Monte Carlo and Brownian dynamics simulations predict results that are in good agreement with each other.  Using kinetic Monte Carlo we find that at high density the particle transport is hindered by jamming. 
\end{abstract}

\maketitle

\section{Introduction}

Manipulation and transport of magnetic particles at the nanometer and micrometer
scales are important technological processes for  biological~\cite{Thomas:2000p3836} and biomedical applications~\cite{Fisher:2005p3838}.
Magnetic microspheres  are routinely used as  markers for cells or larger molecules. 
Micron-sized magnetic beads were attached to the free end of viral DNA~\cite{Chang:2008p3842} for the direct visualization in a bright field microscope of a DNA-packaging process.
\citet{Molday:1977p3843} achieved separation of red blood cells and lymphoid cells using magnetic particles chemically bound to antibodies. \citet{Pamme:2006p3875} achieved continuos cell sorting of magnetically labeled cells via free-flow magnetophoresis.
Magnetic particles can also be tagged with a  fluorescent dye and  followed in real time using microscopy~\cite{Gao:2010p3877}, and their surface can be functionalized to selectively bind to specific targets in a solution, allowing for selective separation using magnetic fields~\cite{kemshead1985}. 

Since a magnetic particle in a homogenous magnetic field cannot have a net translational motion, a variety of methods have been developed for the generation of inhomogeneous magnetic fields that can be used to manipulate and transport magnetic particles. 
\citet{Deng:2001p2879} and \citet{Lee:2001p2883} used lithography to fabricate circuits that carry electrical currents in order to generate the  inhomogeneous magnetic field that is necessary for manipulating the magnetic particles. 
Another approach relies on the deposition of discrete ferromagnetic elements. 
In particular, \citet{Yellen:2003p2334} printed cylindrical magnetic islands on a substrate to drive the assembly of colloidal particles. The application of a rotating external magnetic field allows to control and transport small super-paramagnetic particles~\cite{Yellen:2007p2981,Yellen:2009p2989}. 
\citet{Gunnarsson:2005p2336} deposited elliptical magnetic elements and showed the ability to transport paramagnetic particles along the ellipses by the application of a rotating magnetic field. 
Another technique uses ferrite garnet films~\cite{Dhar:2007}, which show patterns of alternating magnetization, and an  external field that oscillates in time in order to control the transport of colloidal particles~\cite{Dhar:2007,Tierno:2007a,Tierno:2008}. 
The pattern on a garnet film forms spontaneously and is not easily controllable. On the other hand, the  deposition of small magnetic islands~\cite{Yellen:2003p2334, Gunnarsson:2005p2336}, opens up the possibility of creating structured magnetic substrates with full control of the deposition pattern. 

In this light, we carry out both kinetic Metropolis Monte Carlo and Brownian Dynamics computer simulations to study the behavior of paramagnetic colloidal particles on a substrate of discrete magnetic dipoles arranged in two specific patterns, namely parallel and  zigzag stripes. The patterns were inspired by those found in garnet films ~\cite{Dhar:2007,Tierno:2007a,Tierno:2008}, but could, in principle, also be fabricated by one of the above deposition techniques~\cite{Yellen:2003p2334, Gunnarsson:2005p2336}.  The field generated by a garnet film~\cite{Tierno:2007a} is quantitatively different from the one produced by an array of discrete dipoles. Nevertheless, we show that the differences are small and that our model shows a particle transport behavior similar to the one found experimentally on garnet films. 
In particular, we find that the random Brownian motion of the colloidal particles is turned into a deterministic motion by an oscillating external magnetic field. We analyze the conditions that enable the deterministic motion and hence the controlled transport of the paramagnetic colloidal particles.

Brownian dynamics is based on the equations of motion for overdamped particles without hydrodynamic interactions, on the other hand Monte Carlo reproduces the correct dynamics only under certain conditions~\cite{Heyes:1998p3491,Cheng:2006p4039,Kotelyanskii:2011p4040,KIKUCHI:1991p3493,Berthier:2007p2265,Sanz:2010p3489}. 
The advantage of  Monte Carlo is its  higher computational efficiency with respect to Brownian dynamics.
Therefore, we explicitly compare the results of Monte Carlo and Brownian dynamics at low density, and carry out only Monte Carlo simulations in the computationally demanding high density regime.

The paper is organized as follows: In Sec.~\ref{sec:mm}  we summarize the model used and describe the simulation details. In particular, in Sec.~\ref{int} we define the particle-particle and particle-substrate interactions, 
in Sec.~\ref{land} we discuss the energy landscape of the model, and in Sec.~\ref{sim} we give the simulation details.
In Sec.~\ref{sec:res}, we show and discuss the results for both parallel and zigzag stripes. 
In Sec.~\ref{sec:conc} we give some concluding remarks. In appendix~\ref{app} we discuss similarities and differences between the field produced by our model and that produced by a Garnet film.
\begin{figure}[tbp]
\centering \includegraphics[width=9cm]{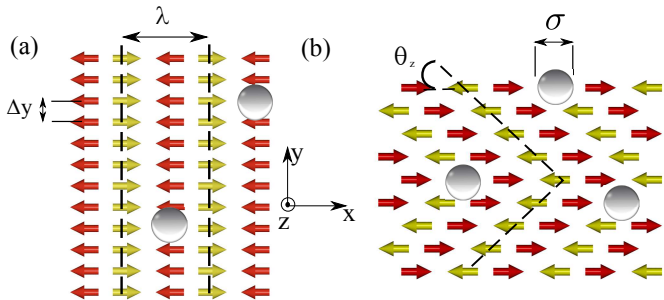}
\caption{(Color online) Sketch of the model. The colloidal particles (spheres) are suspended at a distance $z_{\rm coll}$ over a pattern of discrete point dipoles (arrows). The dipoles are oriented along the $x$-axis with alternating magnetization $\vec{m}=( \pm m_{0} , 0, 0)$. The wavelength of the repeating pattern is $\lambda$. The particles in the figures are shown with diameter $2\sigma$ to help visualization. (a) Parallel stripes pattern. 
(b) Zigzag pattern with  the zigzag angle $\theta_{z}$ as a control parameter. }
\label{fig:model} 
\end{figure}

\section{Model and Method}
\label{sec:mm}

\subsection{Definition of the interactions}
\label{int}
We study a fixed lattice of size $N_{x}\times N_{y}$, of point magnetic dipoles $\vec{m}_{ln}$ lying in the $x$-$y$ plane with components $( \pm m_{0} , 0, 0)$. The lattice sites are enumerated by the pair of integers  $(l,n)$, where $l$ refers to the $x$-direction and  $n$ to the $y$-direction.
The $x$-component  is $+m_{0}$ for dipoles sitting at an odd $l$ position and $-m_{0}$ for dipoles sitting at an even  $l$  position. The dipole moments  have all the same magnitude $m_{0}$ and form a pattern of parallel (Fig.~\ref{fig:model}a) or zigzag stripes (Fig.~\ref{fig:model}b). 
The wavelength of the repeating pattern in the $x$-direction is denoted by $\lambda$. The separation distance between point dipoles in the $y$-direction is $\Delta y$ and the separation distance between point dipoles in the $x$-direction is $\Delta x=\lambda/2$.  The zigzag pattern is characterized  by the angle $\theta_{z}$, as shown in Fig.~\ref{fig:model}b. 

The substrate generates a magnetic field 
\begin{equation}
 \vec{H}_{{\rm sub}}(\vec r)=\sum_{l,n}  \frac{1}{4 \pi } 
 \left(  \frac{ 3 \ \vec{r}_{ln} (\vec{m}_{ln} \cdot  \vec{r}_{ln} )}{ r_{ln} ^{5}}-\frac{\vec{m}_{ln}}{ r_{ln} ^{{3}}} \right) \ ,
\end{equation}
where $\vec{r}_{ln}$ is the distance between the dipole $(l,n)$, and the space point $\vec r=(x,y,z)$. 
In addition, a time-dependent and spatially homogeneous external magnetic field $\vec H_{{\rm ext}}(t)=\vec{H}^{\rm max}_{{\rm ext}} \sin ( 2 \pi t/ \tau_{0})$ is applied to the system. Here $\vec{H}^{\rm max}_{{\rm ext}}=(H^{x},H^{y},H^{z}) $ is the amplitude of the external field, $\tau_{0}$ is its oscillation period, and $t$ is the time.

A  colloidal fluid of $N$ paramagnetic spheres with hard-core diameter $\sigma$ lies suspended at a  distance $z_{{\rm coll}}$ from the patterned substrate and is constrained to move in the $x-y$ plane only.
The total magnetic field exerted on a paramagnetic particle $i$ at position $\vec r_{i}=(x_{i},y_{i},z_{\rm coll})$ is the sum of the external field and the substrate field 
\begin{equation}
\vec{H}(\vec r_{i},t)=\vec{H}_{{\rm sub}}(\vec r_{i}) + \vec{H}_{{\rm ext}}(t)\ .
\label{eqh}
\end{equation} 
Hence, a dipole moment $\vec{m}_{i}=\chi\vec{H}(\vec r_{i},t)$, is induced in the paramagnetic particle $i$ with susceptibility $\chi$. 

The interaction energy between the dipoles in the substrate is constant in time, therefore the relevant energy of our model is the sum of three contributions: first the hard-core interaction between the particles, second the interaction between the particles' (induced) dipole moments $\vec{m}_{i}$ and the total magnetic field $\vec{H}(x,y,z_{\rm coll},t)$, and third the  dipole-dipole interaction between the particles.
The total energy can therefore be written as
\small
\begin{eqnarray} 
&&\beta U_{\rm tot}(t)= \sum_{i <j} \Phi(r_{ij}) -\beta\mu_{s} \chi\sum_{i} \vec{H}(\vec r_{i},t)^{2}   \label{te:tot}  \\
&-& \sum_{i<j}\frac{\beta\mu_{s} \chi^{2}}{4\pi r_{ij}^{3}} \left[ 3 \ \vec{H}( \vec r_{i},t)\cdot\vec{e}_{ij} \ \vec{H}(\vec r_{j},t)\cdot\vec{e}_{ij} -\vec{H}(\vec r_{i},t)\cdot\vec{H}( \vec r_{j},t) \right] \nonumber ,
\end{eqnarray}
 where the hard-sphere potential $\Phi(r_{ij}) =\infty$ if $r_{ij}<\sigma$ and zero otherwise, with $r_{ij}=| \vec r_{i}- \vec r_{j}|$ the distance between colloidal particles $i$ and $j$, $\vec e_{ij}=\vec r_{ij}/r^{2}_{ij}$ , $\vec r_{i}=(x_{i},y_{i},z_{\rm coll}) $ and $\vec r_{j}=(x_{j},y_{j},z_{\rm coll})$.
The field $\vec{H}$ is defined by equation~(\ref{eqh}),  $\mu_{s}$ is  the magnetic permeability of the solvent and  $\beta=1/k_{B}T$, where $k_{B}$ is the Boltzmann constant and $T$ is the temperature. 
 \normalsize
 
\begin{figure}
\includegraphics[width=7cm]{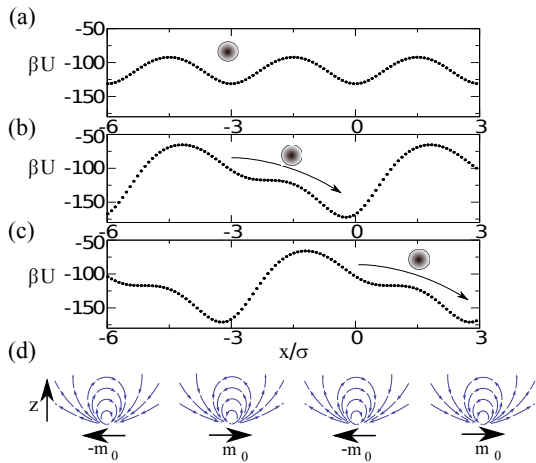}
\caption{(Color online) (a) Scaled energy landscape $\beta U$ as a function of $x/\sigma$ at position $y=0$, and $z=z_{\rm coll}$ for a pattern of parallel stripes at time $t=0$. The circle indicates the preferred position of the particle.
 (b) Same as (a) but at time $t$=$\tau_{0}$/4 in a tilted external potential. The arrow indicates that the particle jumps to the next energy minimum. (c) Same as \rm  (b) but at time $t=3\tau_{0}/4 $.  (d) Magnetic field lines of a sequence of positive and negative point dipoles in the $x-z$ plane.
 These dipoles generate  the energy landscape shown in (a). }
\label{fig:ratpot}
\end{figure}

\subsection{Analysis of the energy landscape}
\label{land}
As shown by Eq. (\ref{te:tot}), the particle-particle interaction  depends quadratically on $\chi$, while the substrate-particle interaction has a linear dependence on $\chi$. Therefore, in the limit of small $\chi$, the particle-substrate interaction is the leading contribution to the total energy. Hence, valuable information about the model can be extracted by simply analyzing the particle-substrate contribution to the total energy, i.e. the limiting case of  a single colloidal particle.
The potential~(\ref{te:tot}) for a single particle, taken as $i=1$, reads 
$\beta U(x_{1},y_{1},z_{\rm coll},t)=-\beta\mu_{s} \chi  \vec{H}(x_{1},y_{1},z_{\rm coll},t)^2$.
Figure~\ref{fig:ratpot}(a) shows this potential as a function of $x=x_{1}$ for the case of $y_{1}=0$ and for a pattern of parallel stripes (as shown in Fig.~\ref{fig:model}a), at time $t=0$. The external field is zero and the energy has a series of minima at the positions of the dipoles (shown in Fig.~\ref{fig:ratpot}(d)), whereas the maximum of the energy is exactly half-way between two dipoles. 
Let us assume that at $t=0$  the particle is sitting in the energy minimum at position $x_{1}=-3 \sigma$, i.e. at the position of a dipole pointing in the positive direction.  
At time $t=\tau_{0}/4 $ the external field has positive $x$- and $z$-components.  Consequently, it reduces the total field above the  dipoles pointing in the positive $x$-direction and enhances the field above the dipoles pointing in the negative  $x$-direction.   Likewise the field between dipoles is enhanced and reduced alternately. 
This gives rise to an asymmetry in the energy landscape, as shown in Fig.~\ref{fig:ratpot}(b). Due to the presence of a point of inflection in the energy, the particle moves towards the energy minimum that is now located above a negative dipole.
At time $t=\tau_{0}/2 $ the external field vanishes, and  the energy landscape is the same  as shown in Fig.~\ref{fig:ratpot}(a). 
On the other hand, at time $t=3\tau_{0}/4$ the external field has negative $x$- and $z$-components. 
The total field above the positive dipoles is now enhanced and the total field above of the negative dipoles is reduced. The field between dipoles is again reduced and enhanced alternately. 
The point of inflection  is again present in the energy landscape (Fig.~\ref{fig:ratpot}(c)) and the particle moves to the next energy minimum, which is now located above a positive dipole.
It is clear from the sequence of Fig.~\ref{fig:ratpot}, that after one cycle $\tau_{0}$ of the external field the particle has covered the distance $\lambda$ and that the cycle can be repeated indefinitely.
When the inclination angle of the external field is such that
either the  $z$-component or the $x$-component is zero, the point of inflection  in the energy landscape is never formed and the particle does not move in any preferential direction. 
For the point of inflection  to form both an $x$- and a $z$-components of the external field are necessary to break the symmetry for a pattern of parallel stripes. 

The particles can on average advance in discrete steps of 0.5 $\lambda$ every half period $\tau_{0}$.
We will hence quantify the transport by the (time) average current $\langle J_{x}\rangle= \langle \frac{1}{N} \sum_{i} \frac{ x_{i} (t) -x_{i}(t_{0})}{t-t_{0}} \rangle$, where $x_{i}(t)$ and  $x_{i}(t_{0})$ are the positions of the colloidal particle $i$ at time $t$ and initial time $t_{0}$, respectively.
With this definition, the maximum current measurable is   $\langle J_{x}\rangle = \lambda/\tau_{0}$, and
an average current  $\langle J_{x}\rangle <   \lambda/\tau_{0}$ is an indication of a decreased efficiency of the transport mechanism, e.g. due to thermal motion or collisions among particles.
Performing computer simulations, as laid out in the next section, allows us to investigate the values of the external magnetic field that induce a particle current and the effect of the particles' Brownian motion as well as effects due to many particle interactions.

\subsection{Simulation Method}
\label{sim}

The simulation box has a lateral size $L_{x} \times L_{y}= 30 \times 60 \ \sigma^{2}$ and has periodic boundary conditions in the $x$- and $y$-directions.
The substrate lies in the $z=0$ plane and contains 2600 point dipoles with dipole moment $m_{0}=50 H_{0}\sigma^{3}$ with $H_{0}$ the unit of the magnetic field. 
We choose a  wavelength $\lambda =3 \sigma$, and a dipole separation distance $\Delta y = 0.3 \sigma$. 
 The colloidal particles are constrained to move in the $z=z_{\rm coll}$=const. plane and have a susceptibility $\chi=0.4 \,  \sigma^{3}$ arbitrarily chosen such that the linear  term in the energy is the leading term. 
We sampled averages for 100 $\tau_{0}$, after 5 $\tau_{0}$ of 'equilibration' time.
The long range dipole-dipole interactions are treated through the Ewald sum~\cite{Ewald:1921p3722,Grzybowski:2002}. 
In order to speed up the interaction calculations, we pre-compute the field  $\vec{H}_{{\rm sub}}$ due to the substrate on a $400 \times 400 $ grid. During the simulations, the field intensity is obtained by interpolation of the  tabulated values. 
The time is in units of the Brownian time $\tau_{B}=\sigma^{2}/D$, with  $D= k_{B} T/\xi$  the Stokes-Einstein diffusion coefficient of the particles and $\xi$ the friction coefficient  of the solvent.
In our simulations, the hydrodynamic interactions are neglected. Due to solvent hydrodynamics, the diffusion coefficient of particles depends on their distance from the substrate~\cite{Reynolds:1886p4048,Goldman:1967p4045,Zahn:1997p4041,Klein:1998p4047,Diamant:2009p4043,Cui:2004p4042,Benmouna:2003p4044,Benmouna:2002p4046}.
Therefore, simulation carried out at constant distance $z_{\rm coll}$ are characterized by a Brownian time $\tau_{B}$ that depends on $z_{\rm coll}$. However, this has no effects  on our results because they are scaled by the Brownian time. 
Further many-body effects due to hydrodynamics are neglected though.

We carry out  both standard Metropolis Monte Carlo~\cite{Allen1987} (MC) simulations with a small MC displacement  $d=0.01 \sigma$, and Brownian Dynamics (BD) simulations~\cite{Allen1987}.
The relationship between MC  and BD has been extensively studied in the literature. 
Both modified MC schemes~\cite{Heyes:1998p3491,Cheng:2006p4039,Kotelyanskii:2011p4040} as well as standard Metropolis MC simulations~\cite{KIKUCHI:1991p3493,Berthier:2007p2265,Sanz:2010p3489} give dynamical properties that can be in good agreement with the results of BD simulations.  In particular, it was recently shown~\cite{Sanz:2010p3489}  that the dynamical properties  obtained from  Metropolis MC simulations are in good agreement with  those obtained from BD as long as the maximum step size of the MC move, $d$, is small enough and the time scale in MC simulations is obtained according to the relations  $\delta t= a d^{2}/6 \,\tau_{B}$ where $a$ is the average acceptance probability of the MC moves.  The relation was demonstrated for one particle in a arbitrary  one-dimensional potential, and verified explicitly for a many-body system and various three-dimensional potentials.  
Scaling the time with the acceptance probability in equivalent to advancing the MC time only when a move is accepted, this concept was called 'internal clock' by~\citet{royall2007}.
Given the presence of an oscillating external field in our model we explicitly investigate the agreement between MC and BD simulations at low particle density. 

We  carried out simulations for a set of four distances between the colloidal particles and the substrates,  $z_{\rm coll}=1.0, 1.5, 1.7, 2 \, \sigma$.  For $z_{\rm coll}=1.0 \, \sigma$ the current $\langle J_{x}\rangle$ was zero in all cases while the results for  $z_{\rm coll}=1.5, 1.7, 2 \, \sigma$ are discussed in detail below.

\section{Results}
\label{sec:res}

\begin{figure}
\includegraphics[width=7cm]{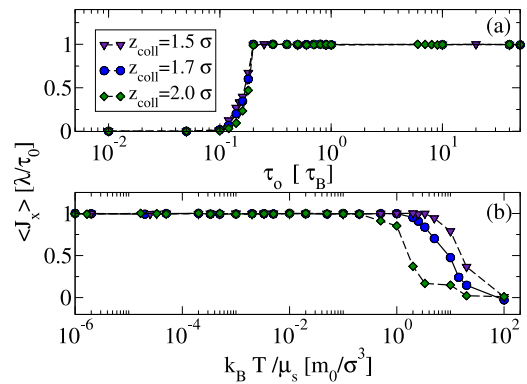}
\caption{(Color online) Average current $\langle J_{x}\rangle$ in the direction perpendicular
to the parallel stripes for $z_{\rm coll}=1.5, 1.7, 2.0 \, \sigma$. (a) As a function of  the  period $\tau_{0}$ of the external
field.  
(b)  As a function of the  temperature-susceptibility ratio, $k_{B}T/\mu_{s}$, with an external field oscillation period $\tau_{0}=10\tau_{B}$.}
\label{fig:ps_temp}
\end{figure}

First we study the low-density behavior of the suspended fluid of colloidal particles on a pattern of parallel lines of  point dipoles, as shown in Fig.~\ref{fig:model}a. We apply a tilted oscillating external magnetic field, with vanishing $y-$component, $H^{y}_{{\rm ext}}(t)=0$, and $x$- and $z$-components given by $H^{x,z}_{{\rm ext}}(t)=H^{x,z} \sin ( 2 \pi t/ \tau_{0})$. The current measurements were conducted using only a single colloidal particle.
Figure~\ref{fig:ps_temp} (a) shows the time average current $\langle J_{x}\rangle$ as a function of the oscillation period of the external field $\vec H_{{\rm ext}}(t)$ for three different values of $z_{\rm coll}$ and with $k_{B}T/\mu_{s}=5 \times 10^{-3}$.
We find that the current is induced only for values of the oscillation period $\tau_{0}$ larger than a critical value. This result can be interpreted easily. If the external magnetic field is oscillating too quickly (small period, high frequency) on the scale of the characteristic diffusion time (Brownian time), then the particle is unable to follow the rapidly changing energy landscape. 
Figure~\ref{fig:ps_temp}(b) shows the $x$-component of the measured current $\langle J_{x}\rangle$,  as a function of   $k_{B}T/\mu_{s}$, for three different values of $z_{\rm coll}$ and with $\tau_{0}=100 \tau_{B}$.

 We find that the external field drives a particle current for values of temperature over permeability, $k_{B}T/\mu_{s}$ smaller than a critical value.  This result suggests that if the thermal energy is too large, the Brownian motion randomizes the motion of the colloidal particles and suppresses the transport.

\begin{figure}
\includegraphics[width=7cm]{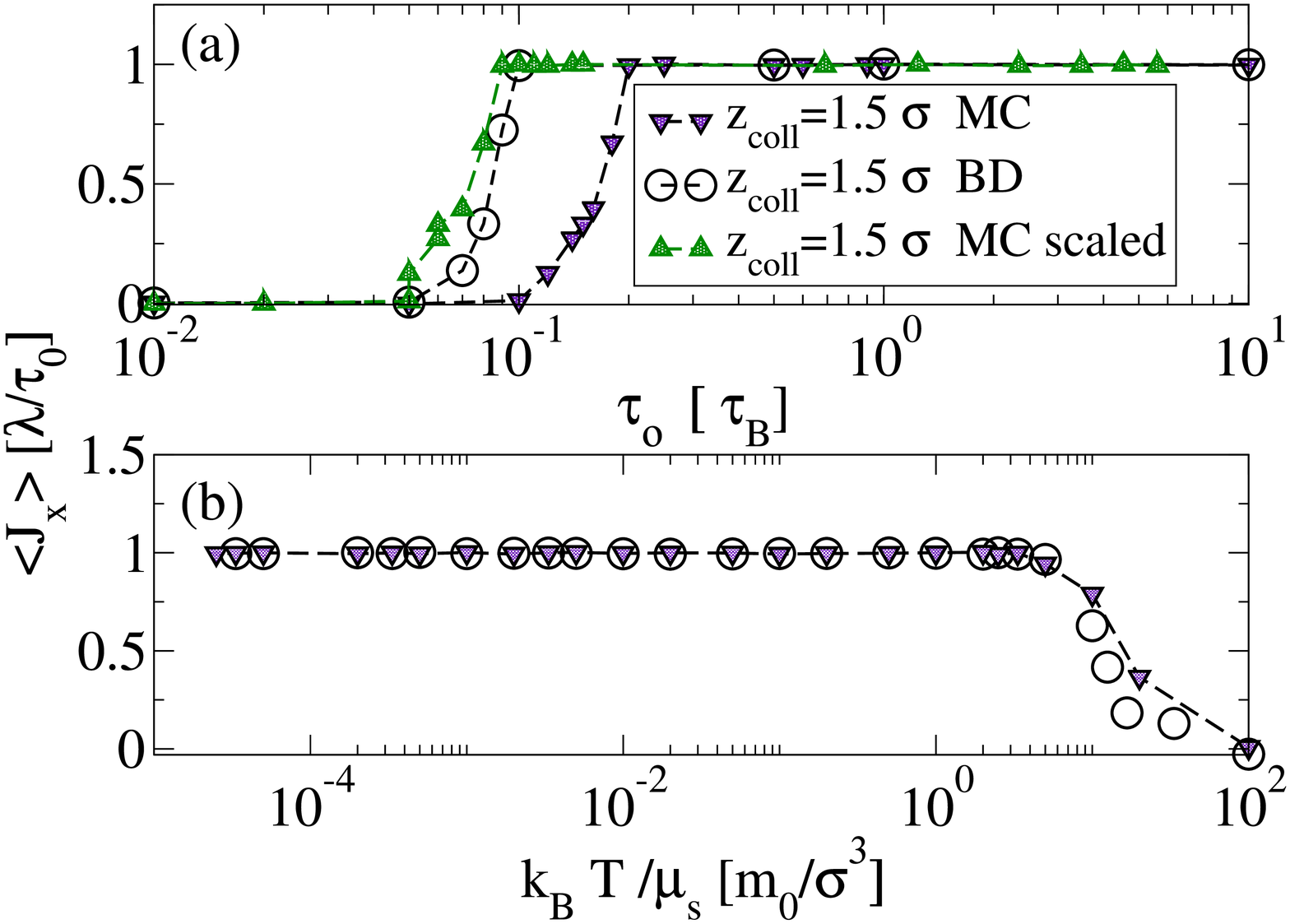}
\caption{(Color online) Comparison between MC, MC scaled with the acceptance probability a, and BD simulation results for the measured average current $\langle J_{x}\rangle$ for $z_{\rm coll}=1.5\, \sigma$. (a) As a function of  the  period $\tau_{0}$ of the external
field.  
(b)  As a function of the  temperature-susceptibility ratio, $k_{B}T/\mu_{s}$, with an external field oscillation period $\tau_{0}=10\tau_{B}$. }
\label{fig:bdmd1}
\end{figure}

Figure~\ref{fig:bdmd1}(a) shows the comparison between MD and BD simulation results for the current as a function of the period of oscillation $\tau_{0}$. The two simulation techniques give qualitatively the same  behavior including the presence of a critical oscillation period beyond which the transport is possible. Nevertheless, the BD predicts a critical period that is two times smaller than that of the MC simulations. We also show the period rescaled by the average acceptance probability  $a$ as suggested by~\citet{Sanz:2010p3489}. The procedure leads to a much better but not perfect comparison.  On the other hand, for oscillation periods far from the critical value the comparison between BD and MD is very good. Figure~\ref{fig:bdmd1}(b) shows that for a value $\tau_{0}=10 \tau_{B}$, both BD and MC simulations give the same behavior for the current as a function of the temperature over permeability.

\begin{figure}
\includegraphics[width=7cm]{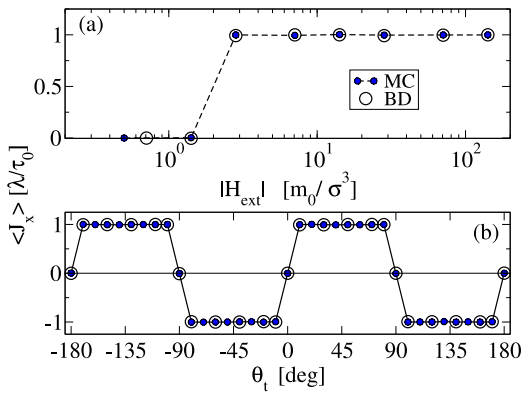}
\caption{(Color online) Average current $\langle J_{x}\rangle$ in the direction perpendicular
to the parallel stripes for  $z_{\rm coll}=1.7 \, \sigma$. (a) As a function of the magnitude $|H_{\rm ext}|$ of the
external field at an inclination angle $\theta_{t}=\arctan(H^{y}/H^{x})=45^{o}$. (b) As a function
of the inclination angle $\theta_{t}$ for a magnitude $|H_{\rm ext}|=50 \ m_{0}/\sigma^{3}$.}
\label{fig:mag}
\end{figure}

Figure~\ref{fig:mag}(a) shows the value of the average current $\langle J_{x}\rangle$ as a function of the magnitude of
the external field $|H_{\rm ext}|$, at a fixed inclination angle $\theta_{t}=\arctan(H^{y}/H^{x})=45^{o}$ of the external magnetic field,  $z_{\rm coll}=1.7 \sigma$ and with an oscillation period $\tau_{0}=10 \tau_{B}$. 
We find that a critical value of the external field needs to be reached in order to initiate the particle transport. 
Figure~\ref{fig:mag}(b) shows the value of the average current $\langle J_{x}\rangle$ as a function
of the inclination angle $\theta_{t}$ at a fixed external field magnitude $|H_{\rm ext}|$ and for  $z_{\rm coll}=1.7 \, \sigma$. We find that the current magnitude and direction can both be controlled by the inclination angle of the external magnetic field.  Note that for angles $\theta_{t}=\pm 180^{o}, \pm 90^{o},0^{o}$ the current is zero. These angles correspond to an external field with either only a component of the external magnetic field parallel to the substrate 
($\theta_{t}=\pm180^{o},0^{o}$) or  only a component normal to the substrate ($\theta_{t}=\pm90^{o}$).  
Figure~\ref{fig:mag} also shows BD simulation results. The comparison between MC and BD simulations is very good.

\begin{figure}
\includegraphics[width=7cm]{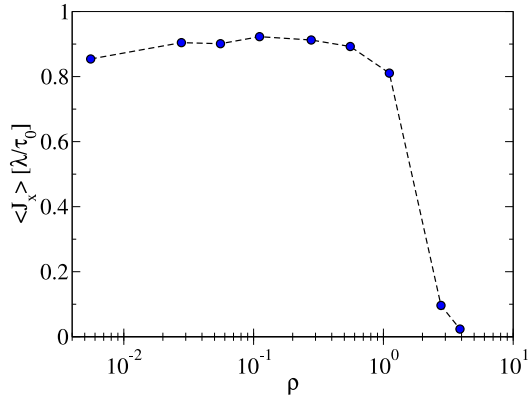}
\caption{(Color online) Average current $\langle J_{x}\rangle$ for a pattern of parallel stripes,  $z_{\rm coll}=1.7 \, \sigma$, $\tau_{0}=10 \tau_{B}$ and  $k_{B}T/\mu_{s}=0.05$  as a function of the linear density $\rho$.}
\label{fig:rho}
\end{figure}

We next investigate the dependence on particle density. We characterize the system by a linear density $\rho=2 \sigma N/N_{x} L_{y}$, where $N$ is the total number of particles. 
Here, we carried out  only MC simulations. These posses higher computational efficiency over BD at the large number of particles that we are considering. 
Figure~\ref{fig:rho} shows the current induced on a parallel stripes pattern, as a function of  $\rho$. 
We find that the current remains roughly constant at low densities with a value slightly smaller than unity. This result indicates a decreased efficiency of the transport mechanism due to the presence of other particles.
Interestingly, the current decays to zero at densities larger than $\rho \simeq 2$. At this density, all stripes are filled with particles. If more particles are present, they must fill positions that are not ideal for the transport mechanism.
These excess particles effectively jam the transport.

\begin{figure}
\includegraphics[width=7cm]{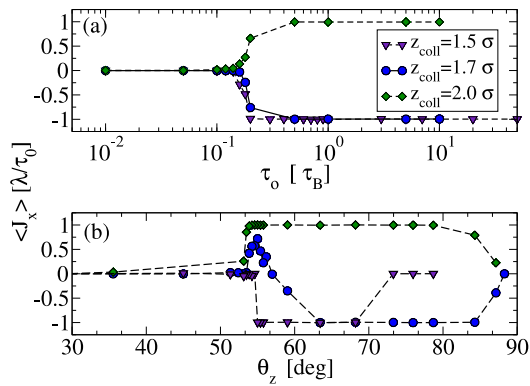}
\caption{(Color online) Average current $\langle J_{x}\rangle$ for a zigzag pattern with $k_{B}T/\mu_{s}=5 \times 10^{-3} $ for  $z_{\rm coll}=1.5, 1.7, 2.0 \sigma$. 
(a) As a function of the oscillation period $\tau_{0}$ of the external field for $\theta_{z}=68^o$. 
(b) As a function of the zigzag angle $\theta_{z}$ for  $\tau_{0}=7 \tau_{B}$.}
\label{fig:slope}
\end{figure}

\begin{figure}
\includegraphics[width=7cm]{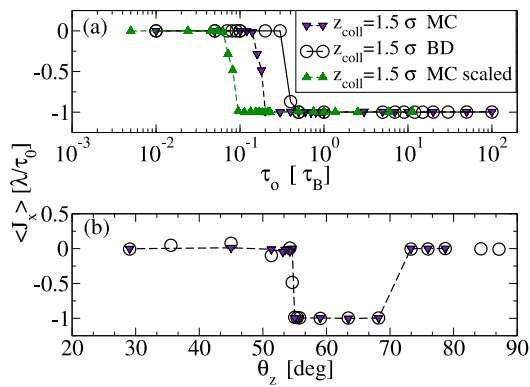}
\caption{(Color online) Comparison between MC, MC scaled with the acceptance probability a, and BD simulation for the measured average current $\langle J_{x}\rangle$ for $z_{\rm coll}=1.5\, \sigma$. a) As a function of the oscillation period $\tau_{0}$ of the external field for $\theta_{z}=68^o$. 
(b) As a function of the zigzag angle $\theta_{z}$ for  $\tau_{0}=7 \tau_{B}$. }
\label{fig:BDslope}
\end{figure}

We next analyze the motion of the particles suspended over a  zigzag pattern with an external field perpendicular to the substrate. The $x$- and $y$-components of the external field vanish, $H^{x,y}_{{\rm ext}}(t)=0$,  and the perpendicular $z$-component, $H^{z}_{{\rm ext}}(t)=H^{z} \sin ( 2 \pi t/ \tau_{0})$, is oscillating in time with period $\tau_{0}$.
Analysis of the particles trajectories shows that transport is achievable only in the vertex region of the zigzag. Therefore, to reliably measure the current, we initialized the simulations with a single colloid randomly positioned along the $x$ direction but localized at the vertex. 
Figure~\ref{fig:slope}(a) shows the current $\langle J_{x}\rangle$ as a function of the  period of oscillation $\tau_{0}$, for three different values of $z_{\rm coll}$ and for $\theta_{z}=68^{o}$.
As for the case of parallel lines, the period of oscillation of the external field needs to be large enough in order to induce a current. Furthermore, we find that for $z_{\rm coll}=1.5, 1.7 \sigma$ the current is negative, while for $z_{\rm coll}=2.0 \sigma$ the current is positive. 
Figure~\ref{fig:slope}(b) shows the average current as a function of the zigzag angle $\theta_{z}$ for external field $H_{z}=10$ $m_{0}/\sigma^{3}$ and $\tau_{0}=100 \tau_{B}$.
The current direction is different  for different distances $z_{\rm coll}$.
In particular it is positive for $z_{\rm coll}=2.0 \sigma$, and negative for  $z_{\rm coll}=1.5 \sigma$. Interestingly, for the intermediate value  $z_{\rm coll}=1.7 \sigma$,  the current is negative for large angles, but positive at smaller zigzag angles.  The current vanishes for $\theta_{z}=90^{o}$, corresponding to the case of parallel stripes.
We stress that the current is only obtained at the zigzag vertex because the external field only has a normal component. This means that the local relative position of the dipole located around the vertex plays a fundamental role in the creation of the point of inflection in the energy, that is necessary for the transport mechanism (see Sec.~\ref{land}). 
Nevertheless, if a tilted external field is applied to the zigzag patterns, transport  is achieved also away form the vertex.
Figure~\ref{fig:BDslope}(a) and (b) show the comparison between MC and BD  simulations for $z_{\rm coll}=1.5 \sigma$. We find a discrepancy in the predicted value of the critical period. Contrary to the case of the parallel lines pattern, the BD predicts a critical period that is two time larger than MD. Rescaling with the average acceptance probability leads to a worse comparison. On the other hand, the comparison is very good for the current as a function of the zigzag angle for a value of the oscillation period $\tau_{0}=7 \tau_{B}$ far from the critical value. 

\section{Conclusions}
\label{sec:conc}
We have studied a simple model for the transport of colloidal particles suspended at a fixed distance over a magnetic patterned substrate with  MC  and BD computer simulations 
Magnetic dipoles were distributed in two specific patterns, namely parallel stripes and zigzag stripes.
We analyzed the effect of an oscillating external magnetic field that was applied to the system.

For the case of parallel stripes we found  that the current magnitude and direction was controlled by the tilt angle of the external field and that the effect was reliably obtained in a wide range of ratios between temperature and solvent  permeability. Furthermore, a net current was measured only when the period of oscillation was greater than a critical value.
For the case of zigzag stripes a current was obtained using an oscillating external field normal to the substrate. 
In this case,  transport was only possible in a small  region of the patterned substrate, namely near the vertex of the zigzag. This result opens up the possibility to transport colloidal particles in a very narrow stream.
Furthermore, the current magnitude and direction was found to be controlled by a combination of the zigzag angle  and the distance of the colloids from the substrate.
The comparison between MD and BD is overall qualitatively very good. We find quantitative agreement  for values of the period of oscillation of the external field far from the critical period, while the two simulations techniques  predicts quantitatively different values of the critical period beyond which transport of particles is possible.

The mechanism behind the transport of the colloidal particles is a consequence of the changing energy landscape.
The sum of the oscillating external field and of  the substrate's magnetic field  results in an energy landscape that changes in time. The Brownian motion enables the particles to locally sample the phase space and follow the energy landscape towards the local (in space and time) energy minimum. Colloidal transport is hence achieved when the particles are able to ``follow'' this landscape.  
The mechanism explained here is the same  as the one  described by~\citet{Dhar:2007} as a deterministic ratchet. 
\citet{Yellen:2007p2981} found the same mechanism and describe it as  particles following a traveling wave. That is, the particle is transported by the translating inflection point in the energy landscape.
 We find that transport of the paramagnetic colloidal particles is possible for a large set of model parameters. 
The magnetic patterns can be created by deposition of discrete magnetic islands~\cite{Yellen:2003p2334,Yellen:2007p2981,Yellen:2009p2989}. Despite that the field generated by a garnet film~\cite{Tierno:2007a} is quantitatively different from the one produced by an array of discrete dipoles, the differences are surprisingly small (see appendix~\ref{app}). Therefore, we expect that the behavior for the transport of particles on top of garnet films is similar to the one shown by our model. 

Controlling the deposition pattern means controlling the behavior of the nano or micro magnetic particles. As a consequence lab-on-chip devices with  well defined functions can be envisioned. 
Our model and  method can be easily applied to  different and more complicated patterns, like for example a combination of parallel and zigzag stripes. Other possible extensions of the current work include the study of substrate boundary effects. In this work we applied periodic boundary condition, but it would be interesting to study the transport of colloidal particles on top of finite discrete patterns.

Furthermore, given the range of phenomena shown by two-dimensional colloidal suspensions of paramagnetic particles trapped at a liquid-air interface (see for example~\citet{Ebert:2009} and references therein) it would be interesting to explore in more detail the effect of a patterned magnetic substrate on the phase behavior and the dynamical properties of two-dimensional fluids. 

\begin{figure}
\includegraphics[width=7cm]{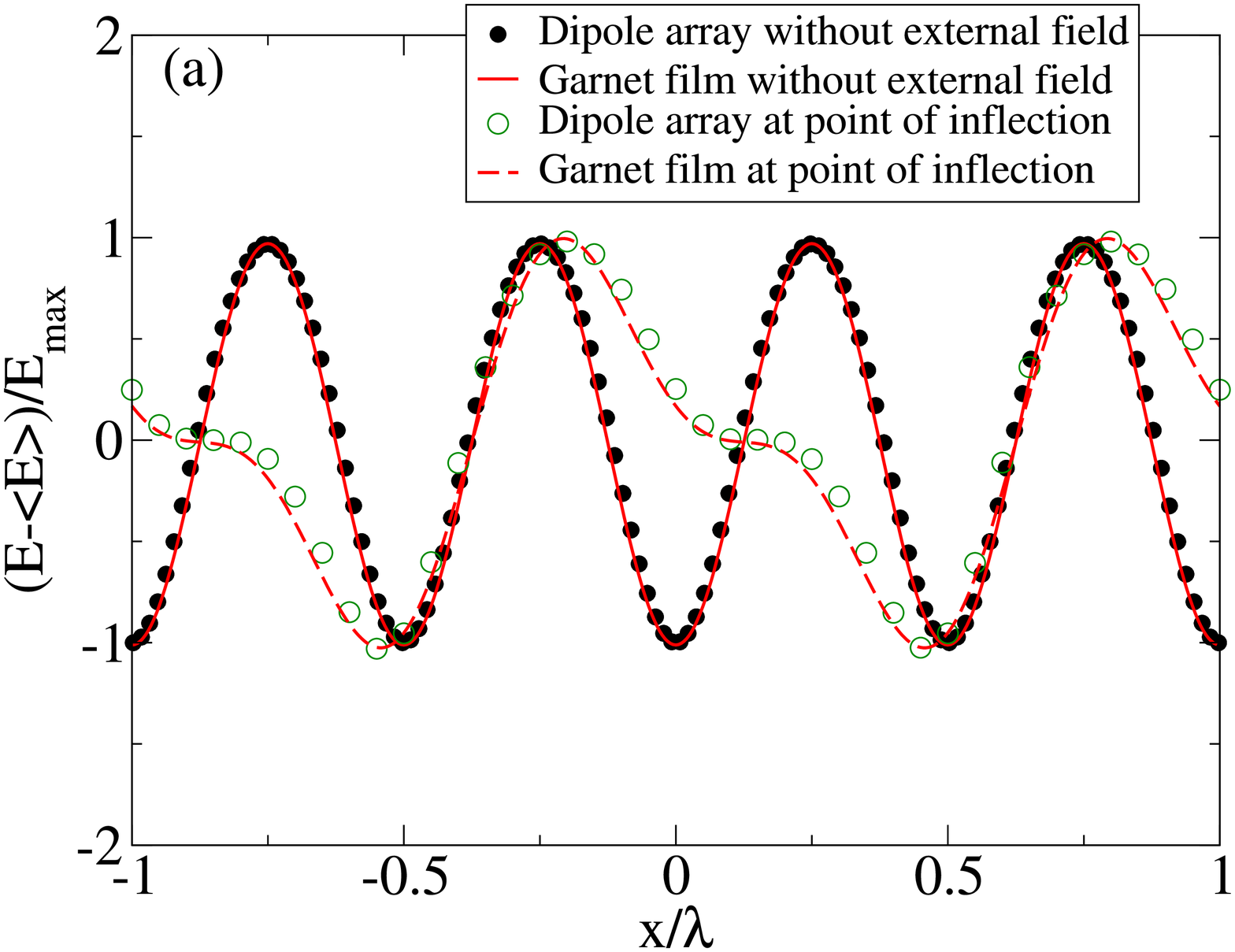}
\includegraphics[width=7cm]{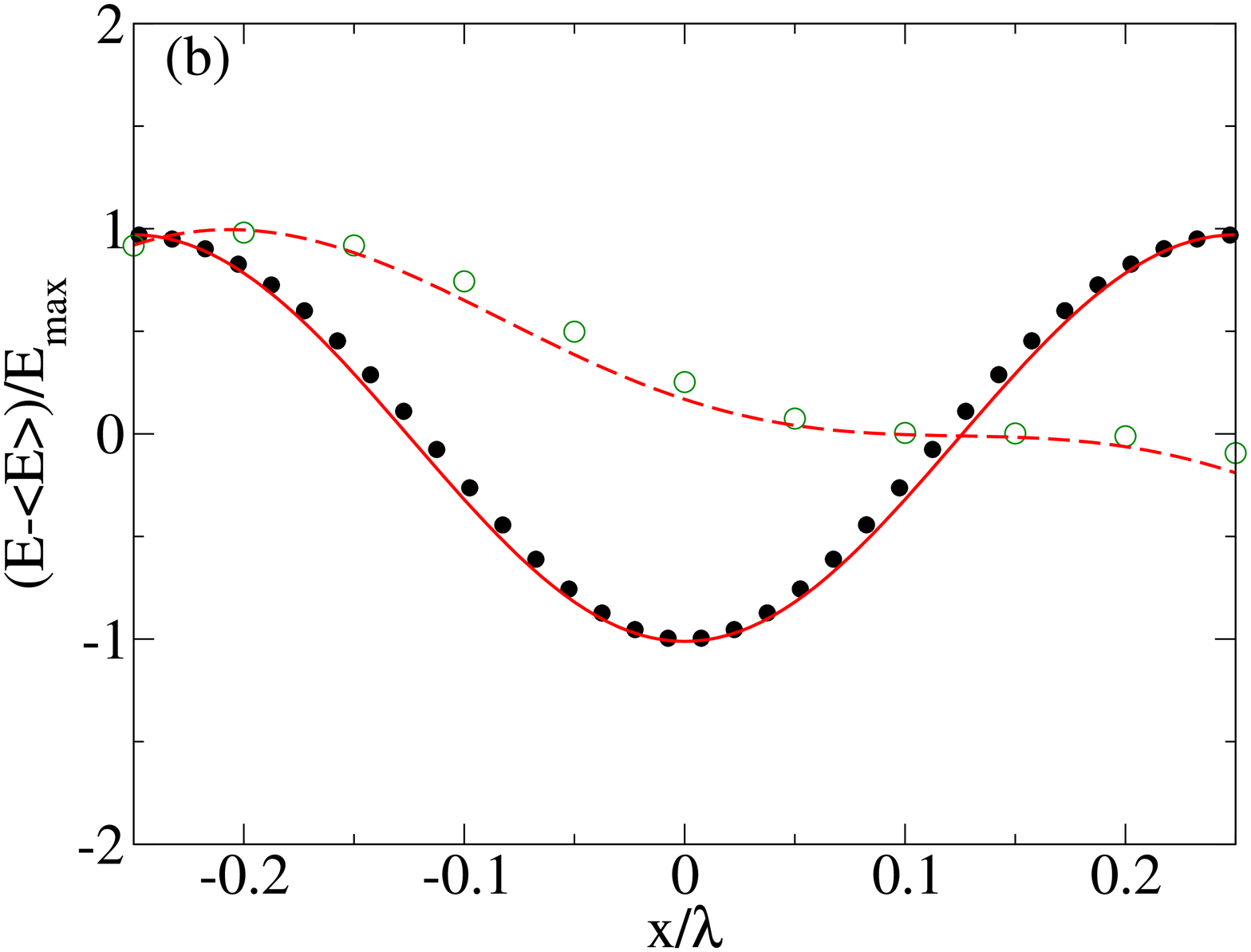}
\caption{(Color online) (a) Comparison between the energy of one paramagnetic particle in the magnetic field of a garnet film and the dipole array. (b) Like (a) but zoomed in to highlight the differences. }
\label{fig:garnet}
\end{figure}

\acknowledgements  We thank Thomas M. Fischer and Saeedeh Aliaskarisohi for useful discussions  and acknowledge the DFG for support via SFB840/A3.

\appendix

\section{Garnet Film}
\label{app}

\citet{Tierno:2007a}  calculated the magnetic field above the garnet film, for a pattern of parallel stripes aligned in the $y$ direction as $\vec H= \vec \nabla {\rm Re}[\Phi]$ with the potential  
\begin{eqnarray}
\Phi &=& \frac{i}{\pi}\text{dilog}(1-\exp(\frac{4 i \pi}{2} w + {\rm Im}(h) )) \nonumber \\
&-&\frac{i}{\pi}\text{dilog}(1+\exp(\frac{4 i \pi}{2}  w -{\rm Im}(h) )) ,
\label{garn}
\end{eqnarray}
 with $w= x+ i z$ and $h=\vec H_{ext} \cdot ( \vec e_{x} - i\vec e_{z})$ and the dilogarithm function $\text{dilog}(t)=\int_{1}^{z} dt \frac{\ln(t)}{(1-t)}$.  Figure~\ref{fig:garnet} shows the comparison between the energy of the garnet film from the potential (\ref{garn}) and the energy of an array of dipoles from equation (\ref{te:tot}).  The energies were shifted by the mean value $\langle E \rangle $ and rescaled by the maximum value of the energy $E_{\rm max}$.
Both with and without external field the differences between the two energies are small, validating the use of the set of discrete dipoles as a good approximation for patterns on garnet films.


\end{document}